\documentclass{aa}

\usepackage{lineno}
\usepackage{hyperref}
\usepackage{siunitx}
\usepackage{booktabs}
\usepackage{tabularx}
\usepackage{array}
\usepackage{graphicx}
\usepackage{subfig}
\usepackage{afterpage}

\usepackage{txfonts}
\usepackage{hyperref}

\begin{document} 

% \linenumbers
   \title{Unraveling the dust activity of naked-eye comet C/2022 E3 (ZTF)}

   \author{Bin Liu
          \inst{1, 2}
          \and
          Xiaodong Liu\inst{1, 2}%\fnmsep\thanks{Just to show the usage
          %of the elements in the author field}
          }
   \institute{School of Aeronautics and Astronautics, Shenzhen Campus of Sun Yat-sen University, Shenzhen, Guangdong 518107, China    \and
             Shenzhen Key Laboratory of Intelligent Microsatellite Constellation, Shenzhen Campus of Sun Yat-sen University, Shenzhen, Guangdong 518107, China \\
             \email{liuxd36@mail.sysu.edu.cn}
             % \thanks{}
             }

   \date{}

% \abstract{}{}{}{}{} 
% 5 {} token are mandatory
 
  \abstract
{A morphological and photometric analysis of the naked-eye long-period comet C/2022 E3 (ZTF) before perihelion is presented in this study. The observation images taken by the Zwicky Transient Facility survey telescope from July 2022 to October 2022 show a gradually brightening dust coma and a tail with a clear structure. The morphology of the dust coma reveals nonsteady-state emission with an ejection velocity lower than 14 m s$^{-1}$ for particles larger than \SI{100}{\um}.
According to the syndyne-synchrone analysis, dust particles larger than about 10 µm contribute significantly to the observed tail.
The model simulations of the 10 October 2022 image suggest that the radii of large particles lingering near the nucleus range from 0.1 mm to 1 mm. Assuming that the nucleus of comet E3 is a homogeneous sphere with an albedo of 0.1, the photometry analysis sets the lower and upper limits of the nucleus radius to be $0.81\pm0.07$ km and $2.79\pm0.01$ km, respectively. The dust production rates increased continuously from $241\pm3$ kg s$^{-1}$ in July to $476\pm9$ kg s$^{-1}$ in October. The dependence of the ejection velocity $v_{\perp}$ perpendicular to the orbital plane of comet E3 on the particle size $a$ can be simplified as $v_{\perp}\propto a^{-1/2}$, which indicates that the dust emission is likely driven by gas. The water-production rate is inferred as $\sim 368\pm72$ kg s$^{-1}$ in October 2022, which is sustained by an equilibrium-sublimating area of $8.2\times10^6$ m$^2$ at least. The comparative analysis of the characteristics of comet E3 with those of comets belonging to different types shows that the activity profile of long-period comet E3 surprisingly aligns more closely with those of short-period comets within a heliocentric distance range of about [1.7, 3.4] AU, where the images of comet E3 that we used in this study were taken.}
   
   
   


   \maketitle
%
%-------------------------------------------------------------------

\section{Introduction}
Long-period comets (LPCs) are classified from a dynamical perspective as comets with orbital periods exceeding 200 years. They are characterized by random orbital inclinations and typically exhibit high eccentricities \citep{lowry2008kuiper}. According to the dynamical evolution model of the early Solar System, LPCs originally formed in the giant planet region (the heliocentric distance ranging from 5 AU to 30 AU) about 4.6 billion years ago \citep{lippi2023volatile}. Subsequently, due to the influence of the orbital migration of giant planets, LPCs were ejected into the outer reaches of the Solar System before they eventually reached their reservoir, known as the Oort cloud, which consists of $10^{12}$ to $10^{13}$ comets with semimajor axes ranging from $10^{3.5}$ AU to $10^{5}$ AU \citep{wiegert1999evolution, nesvorny2018dynamical}. Unlike short-period comets (SPCs), LPCs have experienced minimum heating and sublimation processes since their formation. Their nuclei remain frozen for the majority of their lifetime, suggesting that the material within the nuclei of LPCs has undergone limited alteration since their formation and has largely retained its initial composition \citep{opitom2015trappist}. Hence, the study of LPCs provides a deep understanding of the compositions and conditions of comets at the time of their formation, as well as of the evolution of the protoplanetary disk \citep{hands2020capture}.

LPCs whose orbital paths partially lie within the inner regions of the Solar System undergo continuous heating from solar radiation as they approach perihelion. This process typically leads to the sublimation of volatiles within the nucleus, resulting in the dust activities on the surface of the nucleus \citep{garcia2021photometry}. With the increasing activity of the nucleus, the comet displays the appearance of either a coma or a dust tail \citep{kareta2021activity}. The dust particles in the coma and tail originate from the nucleus, the analysis of which can provide insights into the activity mechanisms and physical properties of LPCs \citep{mumma2011chemical}. However, conducting in situ measurements of LPCs is challenging because the LPC activities are unpredictable \citep{forbes2019turning}, and the current commonly used method for studying the dust activities of LPCs is through observations of specific target areas.

The long-period comet C/2022 E3 (ZTF), hereafter "E3", was discovered on 2 March 2022 with the 48-inch Schmidt-type telescope of the Zwicky Transient Facility (ZTF), with an apparent magnitude of approximately 17.3 at a heliocentric distance of about $R$ = 4.3 AU \citep{bolin2022comet}. The orbit of comet E3 meets the characteristics of Oort-cloud comets, with an eccentricity of 1.0007, an osculating semimajor axis of -4087 AU, and an inclination of 109\degr\footnote{\href{https://ssd.jpl.nasa.gov/tools/sbdb_query.html}{https://ssd.jpl.nasa.gov/tools/sbdb\_query.html}}. Comet E3 reached its perihelion on 12 January 2023 with a heliocentric distance of 1.11 AU. Comet E3 made its closest approach to Earth on 31 January 2023. This observation geometry caused comet E3 to appear quite bright and visible to the naked eye from the Earth, with an apparent magnitude of 5\footnote{\href{http://www.cobs.si/}{http://www.cobs.si/}}. This provides a good opportunity for a detailed characterization of the physical properties of 
comet E3, which is a long-period comet. The closest distance between comet E3 and Earth is only 0.28 AU, indicating that comet E3 is a near-Earth object. Because the warning time during the first approach of E3 to Earth was limited, it is essential to pay special attention to this comet. 

Comet E3 showed cometary features when it was discovered in March 2022 \citep{bolin2022comet}, including 
a green coma, a thin ion tail, and a dispersed yellowish dust tail. As the heliocentric distance of comet E3 decreases with time after its discovery, the solar heat accelerates sublimation of the volatiles on the surface of the nucleus, which produces more gases and increases the brightness of the green coma of comet E3. The appearance of the green coma of comet E3 is probably due to the diatomic carbon ($\mathrm{C}_2$) in the resulting gases by sublimation \citep{borsovszky2021photodissociation,bolin2024palomar,mugrauer2023follow}. The thin ion tail pointed to the antisolar direction, and the yellowish dust tail lay between the antisolar direction and the antivelocity direction of comet E3 \citep{bolin2022comet}. Additionally, an antitail pointing to the Sun became visible when the Earth was close to the orbital plane of comet E3\footnote{\href{http://www.spaceweather.com/}{http://www.spaceweather.com/ (Archive, 22 January 2023})}. Shortly after passing through its perihelion, comet E3 encountered a coronal mass ejection, the resulting solar wind from which broke up its ion tail\footnote{\href{http://www.spaceweather.com/}{http://www.spaceweather.com/ (Archive, 19 January 2023})}. Comet E3 has now moved beyond 4 AU away from the Sun.
%, which was estimated to extend for about millions of kilometers when the comet was near its perihelion \citep{king2022comets}
% These features included a coma, The coma consisted of the gas and the dust ejected from the nucleus of comet E3. 
%The presence of $\mathrm{C}_2$ in comets is common, but it is rare for comets like E3 to be situated so close to both the Sun and Earth and to exhibit such a bright, visible green coma \citep{nationalgeographic2023greencomet}. The other feature that became prominent was its tail-like structure.

The rotation period of the nucleus of E3 was determined to be about 8.49 hours from an analysis of the morphology of the coma in images from the Savonarola Cassegrain telescope, as reported by \citet{manzini2023rotation}. For comparison, \citet{knight2023rotation} reported an apparent period of 8.7 hours using the Lowell Observatory Hall telescope and the Lowell Discovery telescope. Amateur photometric data showed that the brightness of comet E3 continuously increased throughout 2022. The OH production rate was approximately $Q_{\mathrm{OH}} = 1.54\times10^{28}$ s$^{-1}$ on 17 October 2022, as obtained from the TRAPPIST robotic telescopes \citep{jehin2011trappist}, and it increased to $Q_{\mathrm{OH}} = 3.51\times10^{28}$ s$^{-1}$ in the observation taken on 19 December 2022, as the distance from the Sun decreased from 1.96 AU to 1.18 AU \citep{jehin2022trappist, jehin2022trappist1}. The sequence of post-perihelion observation data taken by the Trivandrum Observatory in February 2023 showed a noticeable decrease in the apparent sizes of the head and tail of comet E3 \citep{jayakrishnan2023tracking}. The CN, C$_3$, and C$_2$ production rates on 10 March 2023 were approximately $(5.43\pm0.11)\times10^{25}$ mol/s, $(2.01\pm0.04)\times10^{24}$ mol/s, and $(3.08\pm0.5)\times10^{25}$ mol/s, respectively \citep{bolin2024palomar}.

The European Space Agency (ESA) plans to launch a fast-class (F-class) exploration mission in 2028, named $Comet$ $Interceptor$ (CI) \citep{snodgrass2019european}. CI aims to explore an unknown target, which is classified as a dynamically new comet (a subset of LPCs) \citep{jones2019comet,snodgrass2019european}. The CI spacecraft will initially travel toward the Sun-Earth $L_2$ Lagrangian point and remain there until a suitable target is identified \citep{schwamb2020potential}. The CI spacecraft are planned to be equipped with a dust analyzer, which will collect dust grains from the coma and tail of the comet \citep{moore2021bridge}. Based on the physical properties of LPCs, it is expected that the currently unknown target comet is likely to be in a highly active state. Thus, in preparation for the mission, it is necessary to investigate the dust activity of celestial bodies such as LPCs. Although comet E3 is moving away from the inner Solar System, it is considered to be "the most promising virtual target for CI" \citep{CometInterceptorE3}. Hence, the analysis of the dust activity of comet E3 is important for the science and mission planning of CI.

In this study, we present a morphological and photometric analysis of comet E3 in the images recorded in public science archives to enhance the understanding of its pre-perihelion activity. The structure of the paper is as follows. Section \ref{Data} describes the comet E3 archive that is used in the work in detail. Section \ref{Results} presents the results of the morphological and photometric analysis. The dust properties, nucleus size, production rates, and sublimating area are discussed in Section \ref{discussion}. Section \ref{summary} summarizes the conclusions of this study.

\section{Data}
\label{Data}

The publicly available archival data \citep{https://doi.org/10.26131/irsa22} of comet E3 that are analyzed in this paper are from the ZTF ( \citealt{bellm2018zwicky}), which is mounted on the 48-inch Schmidt-type telescope of the Palomar Observatory. The ZTF is equipped with a 600 megapixel cryogenic charge-coupled device camera, which provides an image scale of $1\arcsec$ per pixel and a field of view of about 47 square degrees. All ZTF images of comet E3 have exposure times of 30 s and are made in the $r$ band. The images of comet E3 were taken from July 2022 to October 2022. It should be noted that only the images taken pre-perihelion were used for analysis in this paper because no images taken post-perihelion that showed clear dust coma and tail are available.
For ZTF images taken on different dates, we removed rays across the entire range of the original image \citep{joye2003new}. By interpolating the areas of the image that were affected by rays with the average pixel count of the surrounding regions, we obtained a cleared image that is free from the effects of the rays \citep{bertin1996sextractor}. The process of data reduction was completed with the Astroart software \citep{nicolini2003astroart}. The details of the observing geometries are listed in Table \ref{geometry}.

\begin{table*}
    \centering
    \renewcommand{\arraystretch}{1.1}
    \caption{Observation geometries of comet E3}
    \label{geometry}
    \setlength{\tabcolsep}{12pt}
    \begin{tabular}{lcccccccc}
                \toprule
Date & DOY$^a$ & $R^b$ & $\Delta^{c}$ & $\alpha^d$ & PsAng$^e$ & PsAMV$^f$ & $\nu^{g}$ & $\delta^{h}$\\
(UT)  &(days) & (AU) & (AU) & (deg) & (deg) & (deg) & (deg) & (deg)\\
\hline
13 Jul 2022 & 193 & 2.86 & 2.18 & 17.6 & 163.9 & 150.4 & 257.2 & -5.5\\
26 Jul 2022 & 202 & 2.70 & 2.12 & 19.9 & 141.3 & 141.5 & 259.7 & -0.1 \\
06 Aug 2022 & 217 & 2.58 & 2.11 & 22.2 & 125.0 & 134.4 & 262.0 & 4.6 \\
06 Sep 2022 & 248 & 2.22 & 2.20 & 26.3 & 90.5 & 119.7 & 269.9 & 16.1 \\
10 Oct 2022 & 282 & 1.83 & 2.25 & 25.8 & 57.8 & 110.8 & 282.1 & 23.5 \\
\hline
\multicolumn{9}{l}{$^a$ Day of year, 2022/01/01 = 1.}\\
\multicolumn{9}{l}{$^b$ Distance between comet E3 and the Sun.}\\
\multicolumn{9}{l}{$^c$ Distance between comet E3 and the Earth.}\\
\multicolumn{9}{l}{$^d$ Phase angle, Sun-E3-Earth.}\\
\multicolumn{9}{l}{$^e$ Angle between the Sun-E3 direction and the north direction (measured counterclockwise).}\\
\multicolumn{9}{l}{$^f$ Angle between the negative projected heliocentric velocity vector and the north direction (measured counterclockwise).}\\
\multicolumn{9}{l}{$^g$ True anomaly of comet E3.}\\
\multicolumn{9}{l}{$^h$ Angle of the observer from the orbital plane of comet E3.}\\
         \end{tabular}
    \end{table*}

\section{Results}
\label{Results}
\subsection{Morphology}
\label{morphology}

\begin{figure*}
\centering
        \includegraphics[width=2\columnwidth]{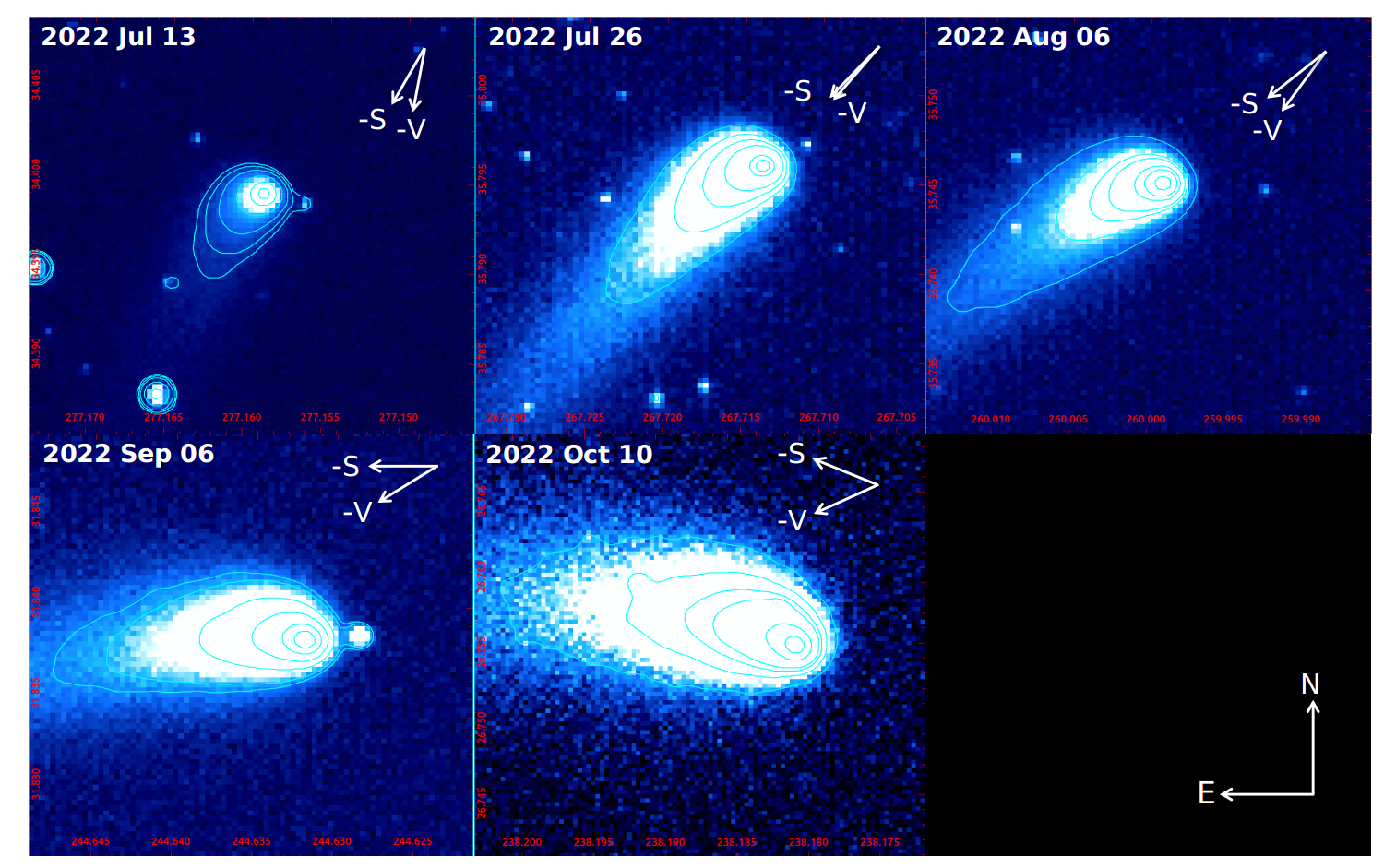}
    \caption{Composite images of comet E3. The observation epoch is marked in the upper left corner of each panel. The projected Sun-E3 radial direction (-S) and the antivelocity direction of the projected motion of comet E3 (-V) are indicated as white arrows. The values of the right ascension and declination are shown in each panel (red text).}
    \label{ObsIma}
\end{figure*}

The images of comet E3 at five different observation epochs are shown in Figure~\ref{ObsIma}. 
% The cometary coma is consistently brightening and exhibits a relative symmetrical appearance.
The coma exhibited a relative symmetrical appearance, and its brightness increased with time. The extended appearance is visible in all observed images. The direction of the tail at different epochs generally lies between the projected direction from the Sun to comet E3 and the antivelocity direction of the projected motion of comet E3, which is a distinctive feature of the dust tails. The length of the tail of comet E3 in the image taken on 13 July 2022 was about 80\arcsec (126000 km), and in the image taken on 10 October 2022, it extended to 
%approximately
about 150\arcsec (245000 km).

\begin{figure}
\centering
        \includegraphics[width=1\columnwidth]{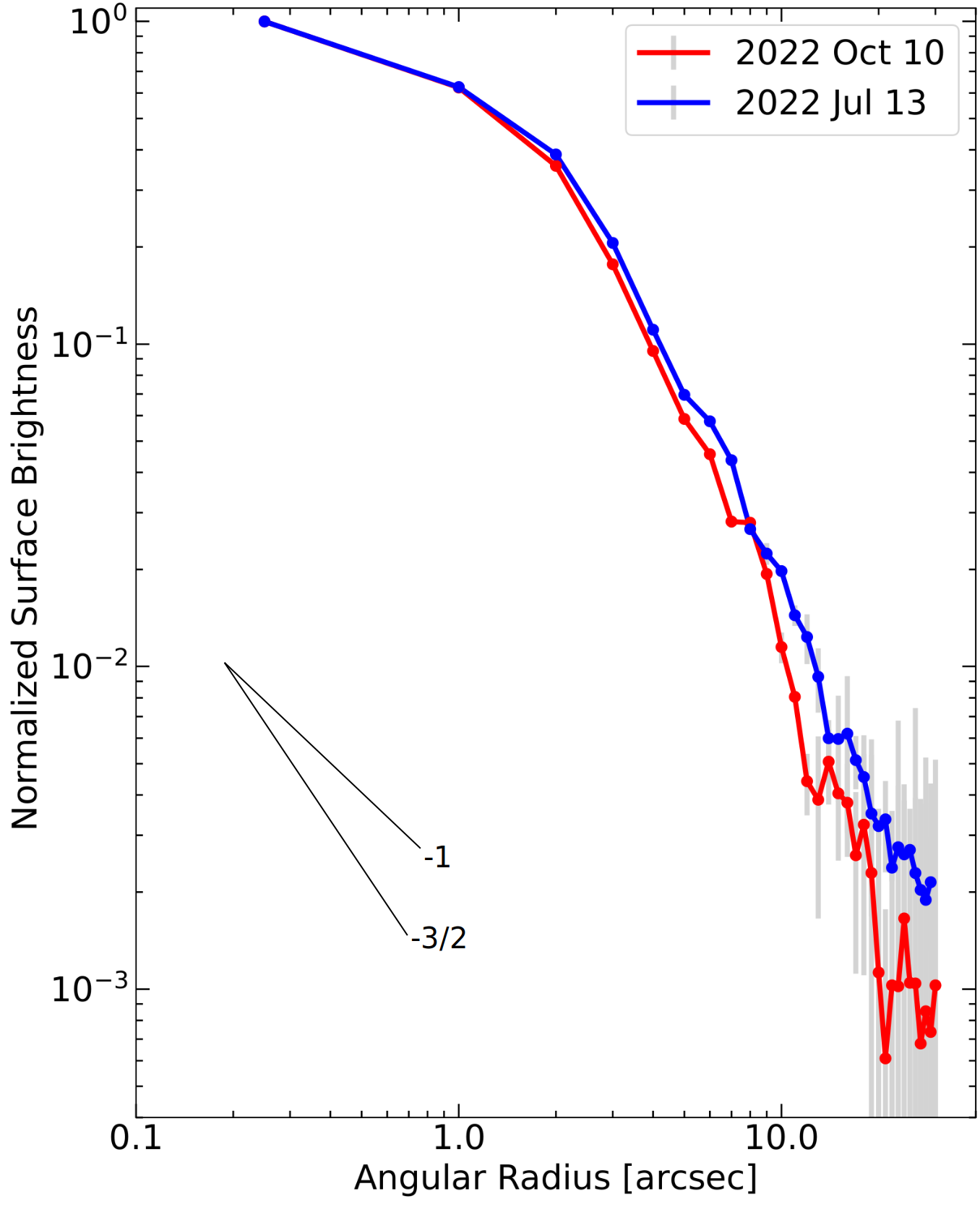}
    \caption{Brightness profiles of the coma measured on 13 July (blue line) and 10 October (red line) 2022. The black lines show the logarithmic gradients of $q = -3/2$ and $q = -1$.}
    \label{ComaSur}
\end{figure}

The surface brightness profiles were determined within a series of concentric annular apertures measured from the optocenter of the coma. Each single annular aperture has a width of 1\arcsec, extending to a radius of 30\arcsec. The sky background was determined in an annulus aperture with an inner radius of 80\arcsec and an outer radius of 120\arcsec. The brightness profiles and their uncertainties resulting from background subtraction for the first and last observed images (13 July and 10 October 2022) are shown in Figure~\ref{ComaSur}. The distribution of the surface brightness $\sum(\theta)$ is well fit to a power-law relation of $\sum(\theta)\propto \theta^{q}$, with $q = -1.61\pm 0.02$ on 13 July 2022 and $q = -1.94\pm 0.03$ on 10 October 2022, where $\theta$ is the angular distance measured from the nucleus in arcseconds. The index $q$ can describe the dynamical state of the coma. For a steady-state coma, the range of the index $q$ is between -1 and -1.5, where the limiting cases of $q = -1$ and $q = -1.5$ correspond to the cases of the absence and the dominance of the solar radiation pressure, respectively \citep{jewitt1987surface}.
% The value of the index $q$ is -1 for a steady-state coma without the effect of solar radiation pressure, while for a steady-state coma for which solar radiation pressure is dominant, the index $q$ is estimated as $-1.5$. 
The obtained indices for comet E3 strongly deviate from the typical value of a steady-state coma, which may suggest changes in the dust \citep{farnham2009coma}.
These deviations are likely caused by the existence of the nonsteady-state emission. The surface brightness profiles were also determined for the images taken on 26 July 2022, 6 August 2022, and 6 September 2022, with indices of $-1.65\pm 0.02$, $-1.74\pm 0.03$, and $-1.81\pm 0.03$, respectively. As the heliocentric distance of comet E3 (Table \ref{geometry}) decreased, the index $q$ increased, suggesting a potential correlation between the heliocentric distance and the index. The research of this correlation relies on the detailed modeling of the dust coma, which is beyond the scope of the present study.

Following Equation (4) from \citet{jewitt1987surface}, the ejection speed of particles $v_{\mathrm{ej}}$ can be estimated by
\begin{equation}
v_{\mathrm{ej}}=\frac{\sqrt{2 \beta \mu_{\odot} \Delta \tan l_{\text {coma }} \sin \alpha}}{R},
        \label{vej}
\end{equation}
where $l_{\text{coma}}\sim 10\arcsec$ ($1.5\times10^7$ m) is the length of the dust coma extending toward the Sun, $\mu_{\odot} = 3.96\times10^{-14}$ AU$^{3}$ s$^{-2}$ represents the gravitational constant of the Sun, and $\beta = C_{\mathrm{pr}} Q_{\mathrm{pr}}/(2 \rho a)$ is a dimensionless factor that represents the ratio of the solar radiation pressure relative to the solar gravity acting on the dust particles \citep{burns1979radiation,moreno2017dust}. Here, $a$ is the particle radius, $\rho$ is the particle bulk density, which is assumed to be $\rho = 500\, \text{kg/m}^3$ \citep{groussin2019thermal}, and
$C_{\mathrm{pr}}$ and $Q_{\mathrm{pr}}$ are the solar radiation pressure parameters, the values of which were adopted as $1.19\times10^{-3}$ m s$^{-2}$ and 1, respectively \citep{burns1979radiation,moreno2017dust}.
Substituting $R, \Delta$, and $\alpha$ of different epochs (Table \ref{geometry}) in Equation \ref{vej}, we estimated the ejection speed as $v_{\mathrm{ej}}\sim139\beta^{1/2}$ m s$^{-1}$. For instance, the value of ejection speed is approximately 14 m s$^{-1}$ for particles with a radius of \SI{100}{\um}. 

\subsection{Photometry}
\label{photometry}

\begin{figure}
\centering      \includegraphics[width=1\columnwidth]{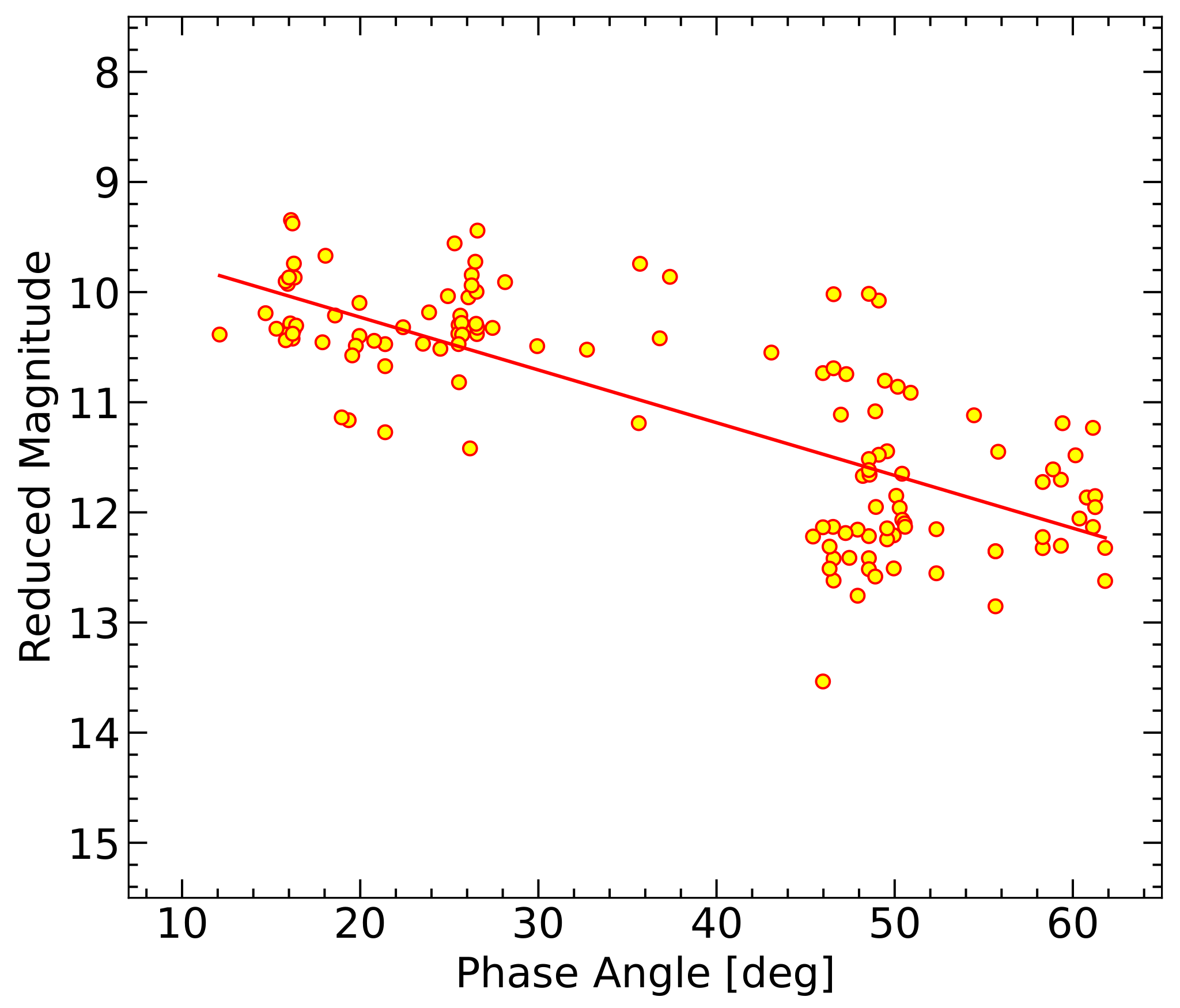}
    \caption{Reduced magnitude vs.~phase angle of comet E3 from 22 March 2022 to 26 March 2023. The filled yellow circles denote the observation data obtained from $CometasObs$. The red line represents the fitting of the data with a slope of ($0.048\pm0.003$) mag deg$^{-1}$.}
    \label{PhaseAng}
\end{figure}

A quantitative analysis is presented for the data of comet E3 at five different observation epochs, using five different sizes of circular photometry apertures with projected radii ranging from 1000 km to 160000 km. The sky background and its uncertainty were also measured in an annulus aperture with an inner radius of 80\arcsec and an outer radius of 120\arcsec as in Section \ref{morphology}. To obtain the absolute magnitudes $H$ in $r$ band, the apparent magnitudes $m_r(R, \Delta, \alpha)$ were reduced to the magnitude at a phase angle of zero degrees and the heliocentric and geocentric distances of one 
AU by equation
\begin{equation}
H=m_r(R, \Delta, \alpha)-5 \log _{10}\left(R \Delta\right)-f(\alpha),
        \label{magnitude}
\end{equation}
where $R$ and $\Delta$ are defined in Table \ref{geometry}. The phase function $f(\alpha)$ can be assumed as a linear function $f(\alpha) = \gamma \alpha$ \citep{delahodde2001detailed}. Because only a limited sample size is available in the ZTF dataset, the linear coefficient $\gamma$ was derived from the amateur photometric data observed by $CometasObs$\footnote{\href{http://www.astrosurf.com/Cometas-Obs}{http://www.astrosurf.com/Cometas-Obs}} covering the period from 22 March 2022 to 26 March 2023.
The magnitude $m$ was reduced to the magnitude at the geocentric and heliocentric distances of one AU \citep{moreno2012comet}, and the relation between the reduced magnitudes $m$ and the phase angle $\alpha$ is shown in Figure~\ref{PhaseAng}. Figure~\ref{PhaseAng} shows that the brightness roughly increases as the phase angle decreases, although the brightness may not solely depend on the phase angle because variations in the activity of E3 at different phase angles might also be a contributing factor. The value of the linear coefficient $\gamma$, ($0.048\pm0.003$) mag deg$^{-1}$ was obtained by fitting the data in Figure~\ref{PhaseAng}, and the phase function $f(\alpha)$ can be simplified and expressed as 0.048$\alpha$. 

From \citet{jewitt2019distant}, the effective scattering cross section, $C_e$ [km$^2$], was derived from the resulting absolute magnitude,
\begin{equation}
C_{e}=\frac{1.5 \times 10^{6}}{p} 10^{-0.4 H},
        \label{cross-secction}
\end{equation}
where $p$ is the geometric albedo. For cometary dust, it is appropriate to assume $p = 0.1$ \citep{ivanova2023quasi}. The values of the resulting apparent magnitudes, the absolute magnitudes, and the effective scattering cross section and their uncertainties for all epochs of observations are listed in Table \ref{PhotoTable}. The uncertainty in Table \ref{PhotoTable} includes only the errors from photometric measurements and does not account for the errors introduced by the determination of the phase function.

\begin{table*}
    \centering
    \renewcommand{\arraystretch}{1.4}
    \caption{Photometry result within apertures of fixed radii}
    \label{PhotoTable}
    \setlength{\tabcolsep}{8pt}
    \begin{tabular}{lcccccc}
        \toprule
        Date & Property$^a$ & 1000 km & 2000 km & 4000 km & 8000 km & 16000 km \\
        \hline
        13 Jul 2022 & $V$ & $17.25\pm0.03$ & $15.94\pm0.02$ & $14.86\pm0.03$ & $14.21\pm0.05$ & $13.73\pm0.06$ \\
        13 Jul 2022 & $H$ & $12.47\pm0.03$ & $11.15\pm0.02$ & $9.98\pm0.03$ & $9.38\pm0.05$& $8.91\pm0.06$ \\
        13 Jul 2022 & $C_e$ & $153.49\pm3.49$ & $515.34\pm4.74$ & $1399.88\pm86.37$ & $2547.37\pm11.7$ & $3945.41\pm72.02$ \\
        26 Jul 2022 & $V$ & $17.11\pm0.02$ & $15.88\pm0.02$ & $14.71\pm0.04$ & $14.11\pm0.03$ & $13.66\pm0.04$ \\
        26 Jul 2022 & $H$ & $12.37\pm0.02$ & $11.15\pm0.02$ & $9.98\pm0.04$ & $9.38\pm0.03$ & $8.91\pm0.04$ \\
        26 Jul 2022 & $C_e$ & $168.33\pm2.34$ & $518.78\pm8.17$ & $1520.89\pm48.27$ & $2666.21\pm83.41$ & $4109.26\pm163.86$ \\
        06 Aug 2022 & $V$ & $17.07\pm0.04$ & $15.84\pm0.03$ & $14.61\pm0.02$ & $13.96\pm0.04$ & $13.48\pm0.03$ \\
        06 Aug 2022 & $H$ & $12.35\pm0.04$ & $11.13\pm0.03$ & $9.89\pm0.02$ & $9.25 \pm0.04$& $8.78\pm0.03$ \\
        06 Aug 2022 & $C_e$ & $171.09\pm5.09$ & $528.51\pm13.17$ & $1666.27\pm36.37$ & $3002.61\pm117.36$ & $4614.93\pm125.59$ \\
        06 Sep 2022 & $V$ & $16.32\pm0.03$ & $15.29\pm0.02$ & $14.41\pm0.03$& $13.62\pm0.04$ & $13.11\pm0.03$ \\
        06 Sep 2022 & $H$ & $11.66\pm0.03$ & $10.63\pm0.02$ & $9.75\pm0.03$ & $8.96\pm0.04$ & $8.44\pm0.03$ \\
        06 Sep 2022 & $C_e$ & $328.16\pm11.87$ & $839.64\pm15.09$ & $1888.39\pm51.08$ & $3945.41\pm177.17$ & $6253.04\pm114.09$ \\
        10 Oct 2022 & $V$ & $15.61\pm0.05$ & $14.98\pm0.03$ & $13.79\pm0.02$ & $13.07\pm0.03$ & $12.48\pm0.02$ \\
        10 Oct 2022 & $H$ & $11.33\pm0.05$ & $10.71\pm0.03$ & $9.53\pm0.02$ & $8.81\pm0.03$ & $8.21\pm0.02$ \\
         10 Oct 2022 & $C_e$ & $442.37\pm21.18$ & $787.21\pm28.26$ & $2315.99\pm44.66$ & $4526.61\pm159.92$ & $7812.26\pm154.43$ \\
        \hline
        \multicolumn{7}{l}{$^a$ $V$: apparent magnitude, $H$: absolute magnitude, $C_e$: effective scattering cross section in km$^2$.}\\ 
    \end{tabular}
\end{table*}

\section{Discussion}
\label{discussion}
\subsection{Dust properties}
\label{Dust properties}

\begin{figure*}
\centering
    \subfloat{\includegraphics[width=0.49\textwidth]{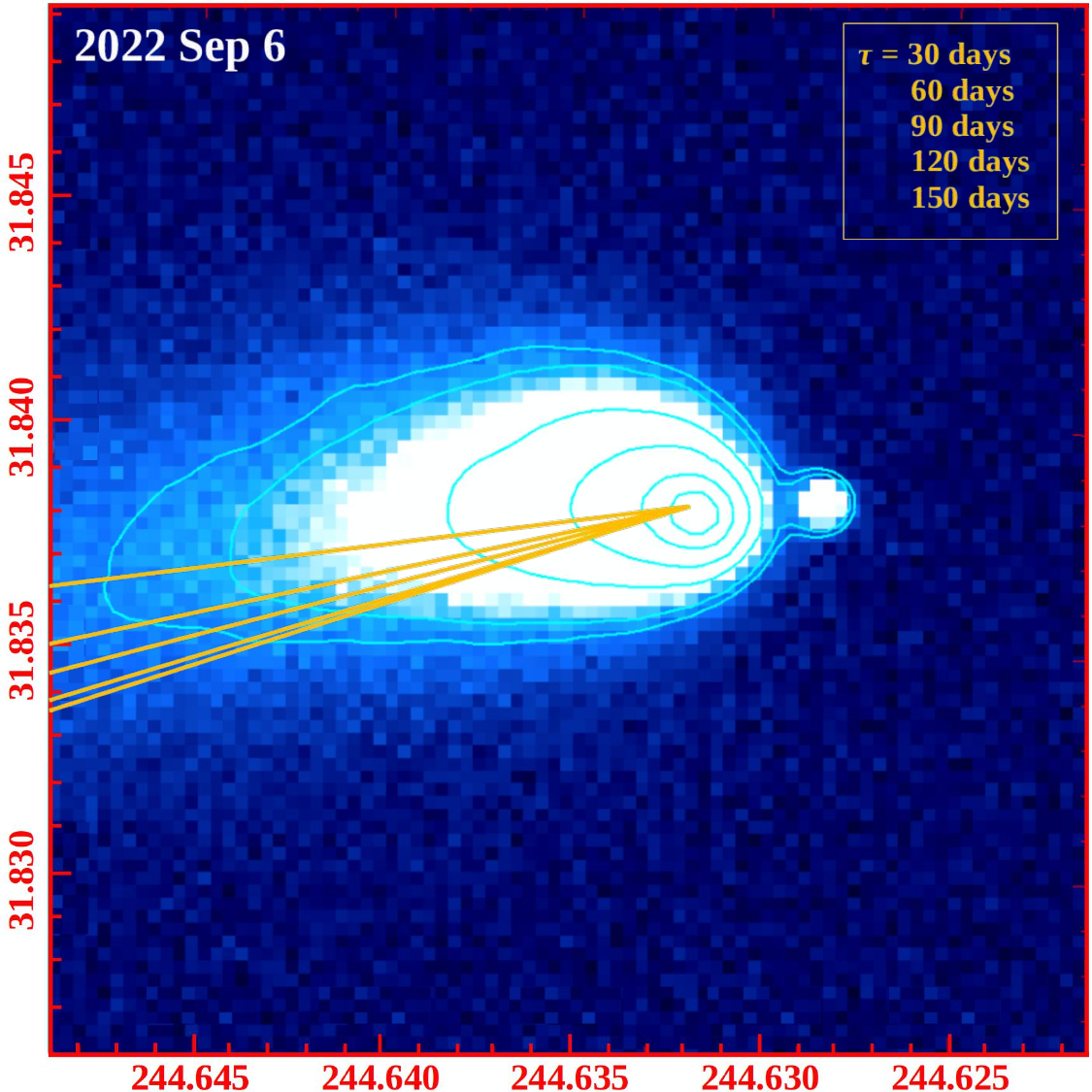}}\hspace{0.001\textwidth}
    \subfloat{\includegraphics[width=0.49\textwidth]{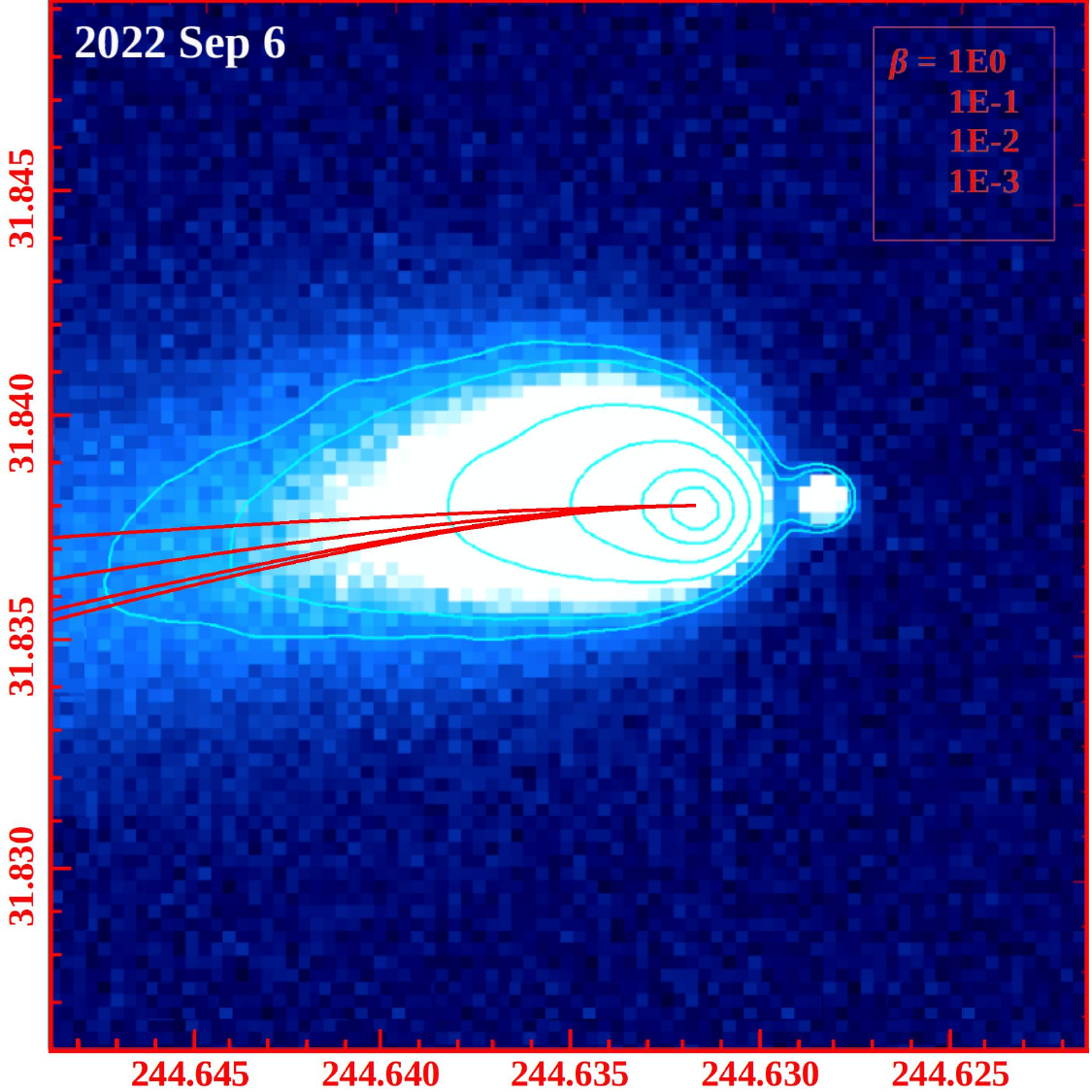}}\\
    \subfloat{\includegraphics[width=0.49\textwidth]{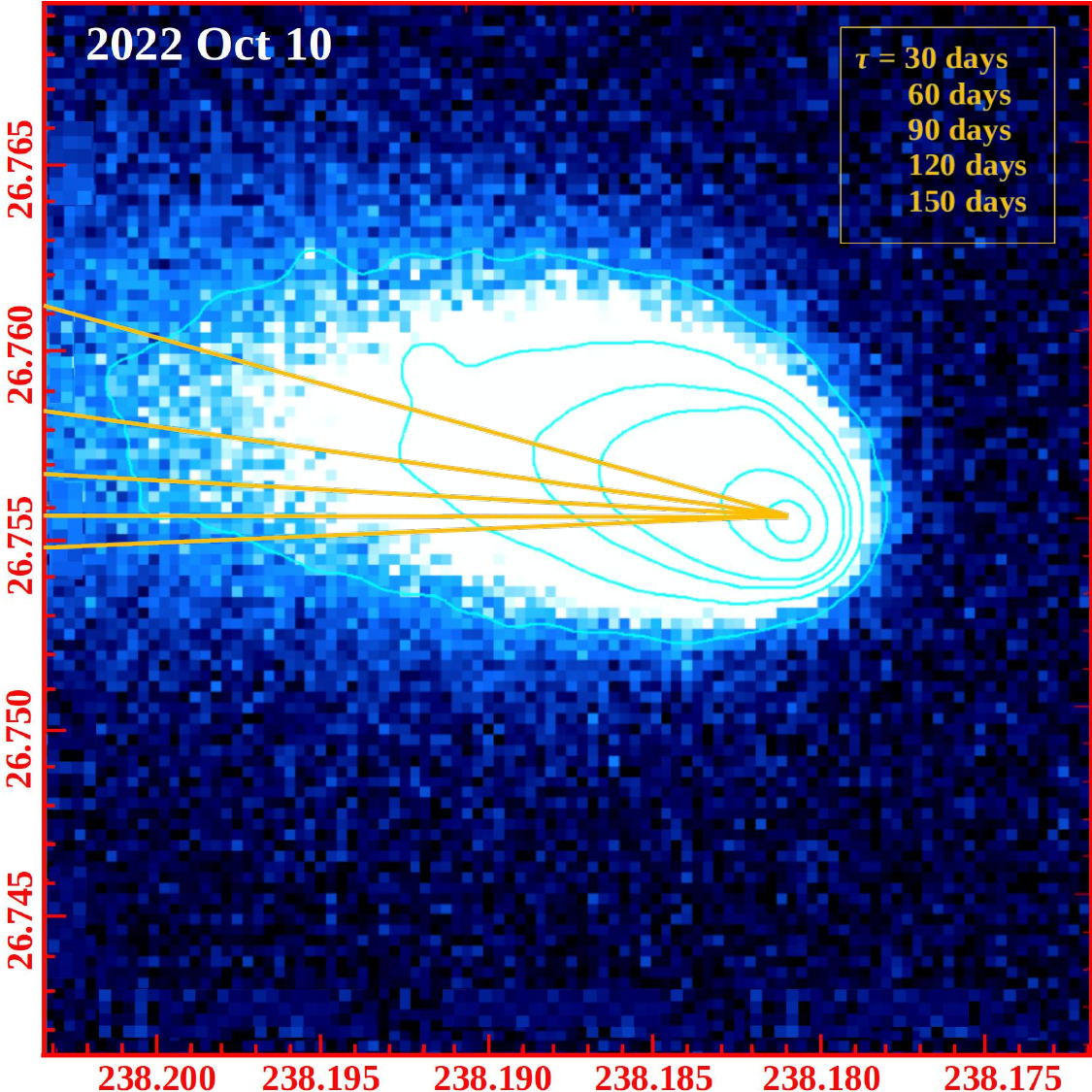}}\hspace{0.001\textwidth}
    \subfloat{\includegraphics[width=0.49\textwidth]{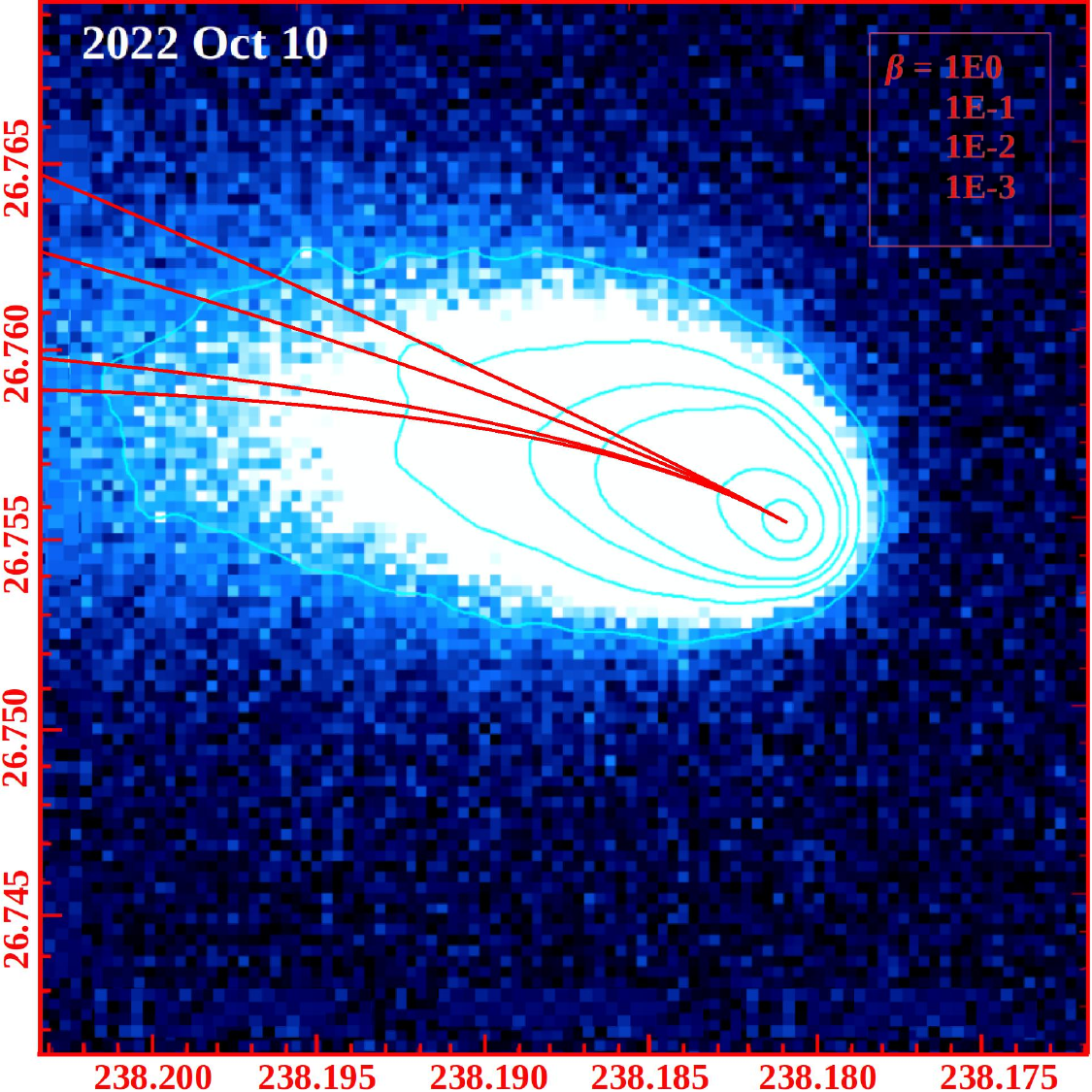}}
    \caption{Syndyne-synchrone grids overplotted on the observation images of comet E3. The yellow lines denote the synchrones of 30, 60, 90, 120, and 150 days (from top to bottom) before the respective observation dates, and the red lines denote the sydynes of $\beta = 1, 1\times10^{-1}, 1\times10^{-2}$, and $1\times10^{-3}$ (from top to bottom). The images are oriented with north up (increasing DEC in degrees) and east to the left (increasing RA in degrees).}
    \label{syncurve}
\end{figure*}

It is possible to determine the physical properties of the cometary grains from the morphology of the coma and tails \citep{kim2022hubble}. The syndyne-synchrone method based on the Finson-Probstein theory \citep{finson1968theory} was employed to analyze the morphology of comet E3. 
The syndyne-synchrone method assumes that only solar gravity and the solar radiation pressure force act on the dust particles. A syndyne corresponds to the loci of the particles that are released continuously with a constant $\beta$. A synchrone line corresponds to loci of the particles that are released at the same time with various $\beta$. For the purpose of model simplification, the ejection speed of the particles was not taken into account when we used the syndyne-synchrone method. 
During all observation dates, the appearances of comet E3 were clear. Thus, it is appropriate to analyze the morphology of comet observed on five different dates using the syndyne-synchrone method with five dense syndyne-synchrone grids. The syndynes were plotted from $\beta = 0.0001$ to $\beta = 1$ with a step of an order of magnitude, and the synchrones are plotted from two years before the observation date up to the observation date with an interval of one day.
In Figure \ref{syncurve}, sparser syndyne-synchrone grids than what we used for the fitting are overplotted on the images for better visualization. Consistent fitting results that can explain all observations were obtained through analyzing the morphology of comet E3 in all images using the syndyne-synchrone grids. The linear jet-like feature is well fit by the synchrone representing about 10 August 2022, and dust particles larger than about 10 µm contribute significantly to the observed tail. The syndyne-synchrone method may no longer be reliable when the upper limit of the grain size is estimated. This is because large particles may linger around the nucleus, which usually cannot be resolved, and they are optically unimportant.
%It is found that the majority of the grains composing the dust tail can be constrained by syndynes with $\beta$ < 0.1 (corresponding to the grain size range of $a$ > $10 \SI{}{\um} = 1/\beta$), indicate that dust particles smaller than about 30 µm contribute very little to the observed tail.

\begin{figure*}
\centering
        \subfloat{\includegraphics[width=0.49\textwidth]{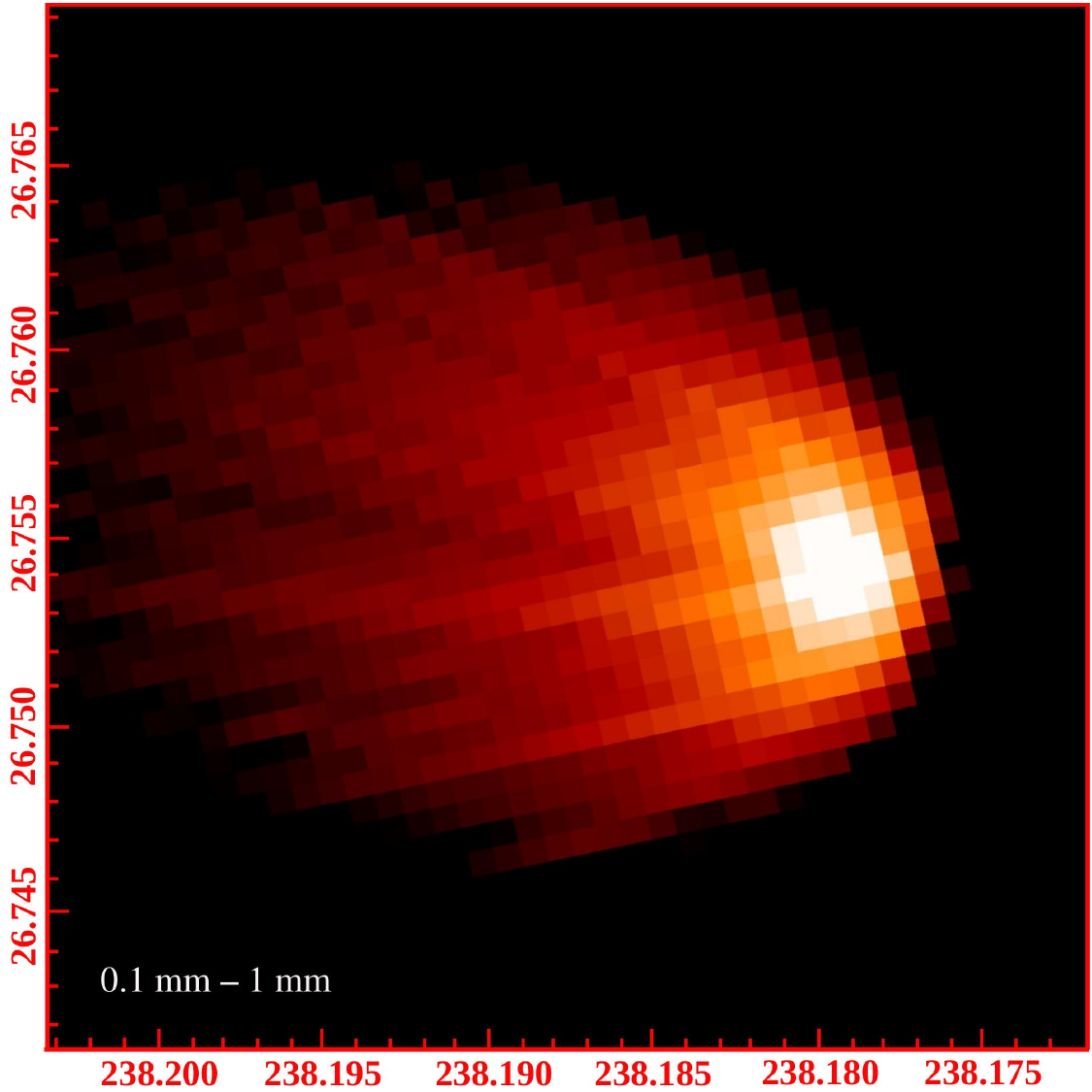}}\hspace{0.001\textwidth}
    \subfloat{\includegraphics[width=0.49\textwidth]{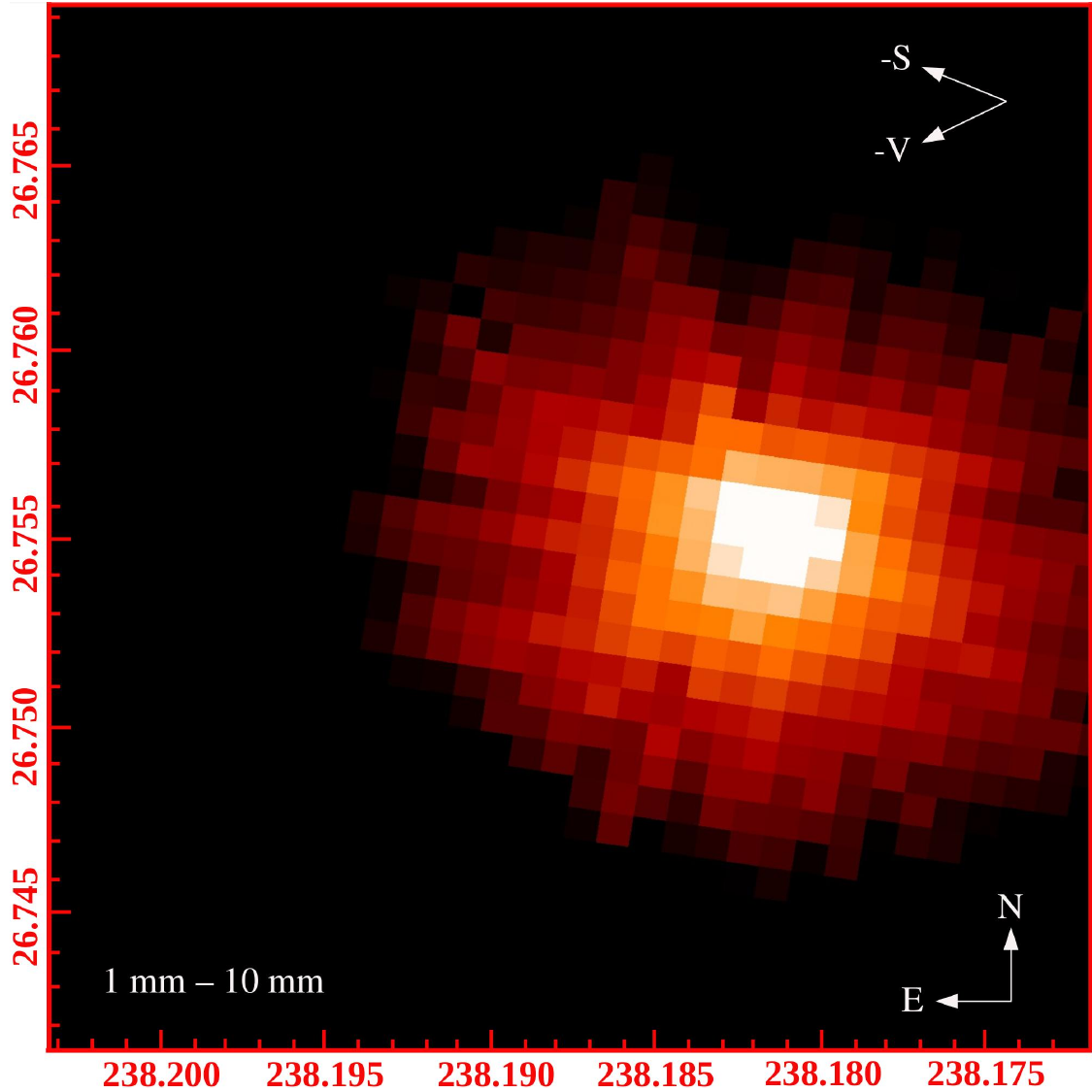}}\\
    \caption{Model simulations of the 10 October 2022 image. The ranges of the dust sizes used for the simulations are marked in the lower left corner of each panel. The projected Sun-E3 radial direction (-S) and the antivelocity direction of the projected motion (-V) of comet E3 are indicated as white arrows. Each panel is oriented with north up (increasing DEC in degrees) and east to the left (increasing RA in degrees).}
    \label{dustmodel}
\end{figure*}

% The large particles contribute significantly to the total mass of the dust tail.
In order to estimate the upper limit of the grain size, we simulated the modeling of the 10 October 2022 image using a dust dynamical procedure. This procedure was developed by \citep{liu2016dynamics} and was modified for this work. In the simulations, the dust ejecta was assumed to be ejected from the sunlit side of comet E3 with a semi-opening angle of $90^\circ$. The starting time of the activity was assumed to be 2 March 2022, which is the epoch when comet E3 was first discovered, and the end time was the observation date. The size distribution of the dust grains followed a differential power-law distribution with an exponent of -3.5. The ejection velocity is related to the particle size and satisfies the relation $v_{\mathrm{ej}}\sim139\beta^{1/2}$ (Section \ref{morphology}). The escape velocity is approximately determined to be 1.47 m/s in the following section \ref{Nucleus size and dust production rates}, and the upper limit size of the escape particles was preliminarily estimated to be 10 mm. Since the morphology of the dust coma on 10 October of 2022 is clearest in Figure \ref{ObsIma}, the simulation was conducted based on this image. The model simulations of the 10 October 2022 image for particle radii of $0.1 \sim 1$ mm and $1 \sim 10$ mm are showed in Figure \ref{dustmodel}. The nearly circular morphology in the image for $1 \sim 10$ mm shows a distribution that is closely isotropic. The image for smaller particles ($0.1 \sim 1$ mm) shows a tadpole-like distribution, which is more closely aligned with the morphology of comet E3. The comparison between the observation and the model simulations of the 10 October 2022 image indicates that large particles lingering near the nucleus of comet E3 are sensitive to radiation pressure. The radii of the majority of these large particles range from 0.1 mm to 1 mm, and the upper limit of the particle size was therefore set to 1 mm.
% \begin{figure*}
% \centering
%       \includegraphics[width=1.5\columnwidth]{Fp_overplot.png}
%     \caption{Syndyne-synchrone grids overplotted on the observation images of comet E3. The synchrones (yellow lines) are for dates of 1 December 2021 and 31 January, 2 April, and 1 June of 2022, and the sydynes (red lines) are for $\beta = 1, 1\times10^{-1}, 1\times10^{-2}$, and $1\times10^{-3}$ for C/2022 E3 (ZTF) on 6 September and 10 October of 2022.} 
%     \label{syncurve}
% \end{figure*}

\subsection{Nucleus size and dust production rates}
\label{Nucleus size and dust production rates}

The effective scattering cross section $C_e$ of the particles within the circular photometry aperture with a projected radius of 1000 km in Table \ref{PhotoTable} can be used to estimate the upper limit of the nucleus radius by $r_n = (C_e/\pi)^{1/2}$.
To impose a stronger constraint on the radius of the nucleus, the aperture photometry of the nucleus was remeasured. A circular aperture with a fixed radius of 1000 km was still used to obtain the photometry of the nucleus. To reduce the dust contamination in the aperture on the nucleus photometry as much as possible, the background of the coma was determined in an annulus aperture with an inner radius of 1000 km and an outer radius of 2000 km. The minimum effective scattering cross sections of the nucleus were derived from the photometry of the image observed on 13 July 2022, giving $C_e = (24.45\pm0.02)$ km$^2$. Assuming that the nucleus of comet E3 is a homogeneous sphere and that the upper limit of size of the nucleus is $r_n = (C_e/\pi)^{1/2}$ = $(2.79\pm0.01)$ km, the upper limit of the mass of a spherical nucleus with an assumed bulk density of
$\rho$ = 500 kg m$^{-3}$ \citep{groussin2019thermal} is $(4.55\pm0.04)\times10^{13}$ kg, and the upper limit of the escape velocity is 1.47 m s$^{-1}$.

\begin{figure}
\centering
        \includegraphics[width=1\columnwidth]{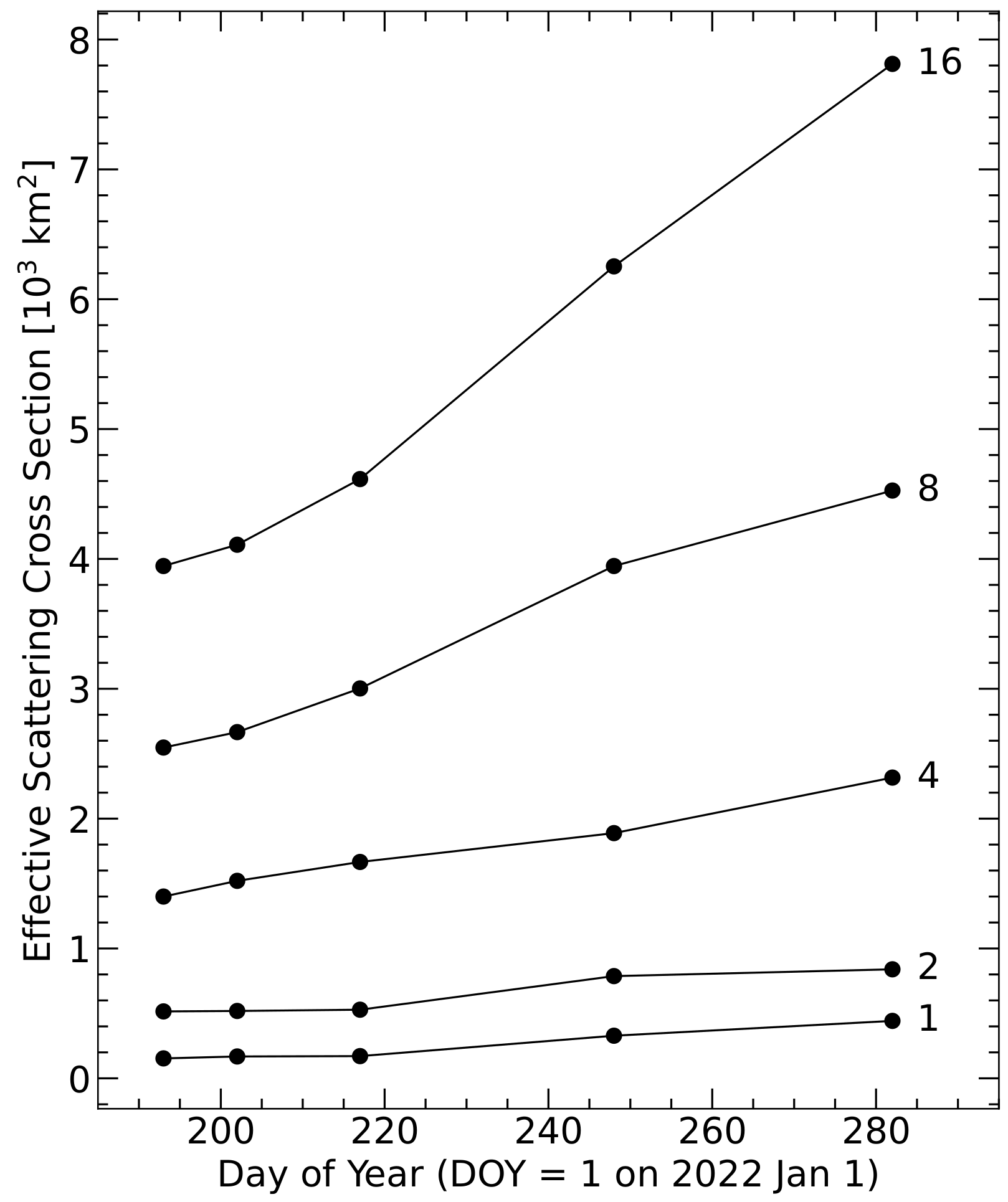}
    \caption{Effective scattering cross section vs.~epoch of observation (day of year = 1 on 1 January 2022) for each of the five apertures with different radii. The radius of each aperture is given at the right side of each line (in units of $10^3$ km). The error bars are smaller than the size of the data points, causing them to be hidden behind the points.}
    \label{CrossSection}
\end{figure}

The variation of $C_e$ (Figure \ref{CrossSection}) can reflect the enhancement or weakening in the activity level of comet E3.
The increase in the variable $C_e$ implies that the inflow of the particles into the aperture exceeds the outflow, and the decrease implies that the outflow exceeds the inflow \citep{jewitt2020outburst}. From \citet{kim2020coma}, the average net dust production rate can be estimated by 
\begin{equation}
\frac{\mathrm{d} M}{\mathrm{d} t}=\frac{4}{3} \frac{\rho \bar{a} C_{e}}{\tau_{r}}.
        \label{ProductionRate}
\end{equation}
Here, $\overline{a}=a_{\mathrm{max}} \times a_{\mathrm{min}}$ is the mean radius of the dust grains, the value of which is estimated as $\overline{a}$ = $\SI{100}{\um}$ (see Section \ref{Dust properties}) under the assumption that the size distribution of the dust grains follows a differential power-law distribution with an exponent of -3.5.
%The size of dust particles is estimated} within the size range of 10 < $a$ < 100 $\SI{}{\um}$} in Section \ref{Dust properties}, which gives the value of $\overline{a}$ = $\SI{30}{\um}$.
The bulk density of the particles is $\rho = 500$ kg m$^{-3}$, as assumed earlier in this section. The values of $C_e$ are seen in Table \ref{PhotoTable}.
The dwelling time of the particles within the aperture is estimated by $\tau_r = L/v_{\mathrm{ej}}$, and an aperture with a radius of $L = 16000$ km was used for photometry. Considering that the ejection velocity of particles with an average radius of about 100 \SI{}{\um} is about $v_{\mathrm{ej}}\sim14$ m s$^{-1}$ according to Equation \ref{vej}, the dwelling time is derived to be about $\tau_r \sim 1.1\times10^6$ s.
%We take $v_{\mathrm{ej}}\sim25$ m s$^{-1}$ with the mean radius of the particles of about 30 \SI{}{\um} (Section \ref{morphology}) to find $\tau_r \sim 6.4\times10^5$ s. 
We substituted $C_e$ within the aperture with a radius of 16000 km (Table \ref{PhotoTable}) in Equation (\ref{ProductionRate}), and we derived the lower limit of the dust production rates as $\sim241\pm3$ kg s$^{-1}$ on 13 July 2022, $251\pm9$ kg s$^{-1}$ on 26 July 2022, $281\pm6$ kg s$^{-1}$ on 6 August 2022, $382\pm6$ kg s$^{-1}$ on 6 September 2022, and $476\pm9$ kg s$^{-1}$ on 10 October 2022. Assuming that the mean radius of the particles remains constant over time, it is found that the dust production rate increases with the decrease of the heliocentric distance (Table \ref{geometry}), suggesting that the activity is highly correlated with temperature. 

\subsection{Plane-crossing}
\label{Plane-crossing}

\begin{figure}
\centering
        \includegraphics[width=1\columnwidth]{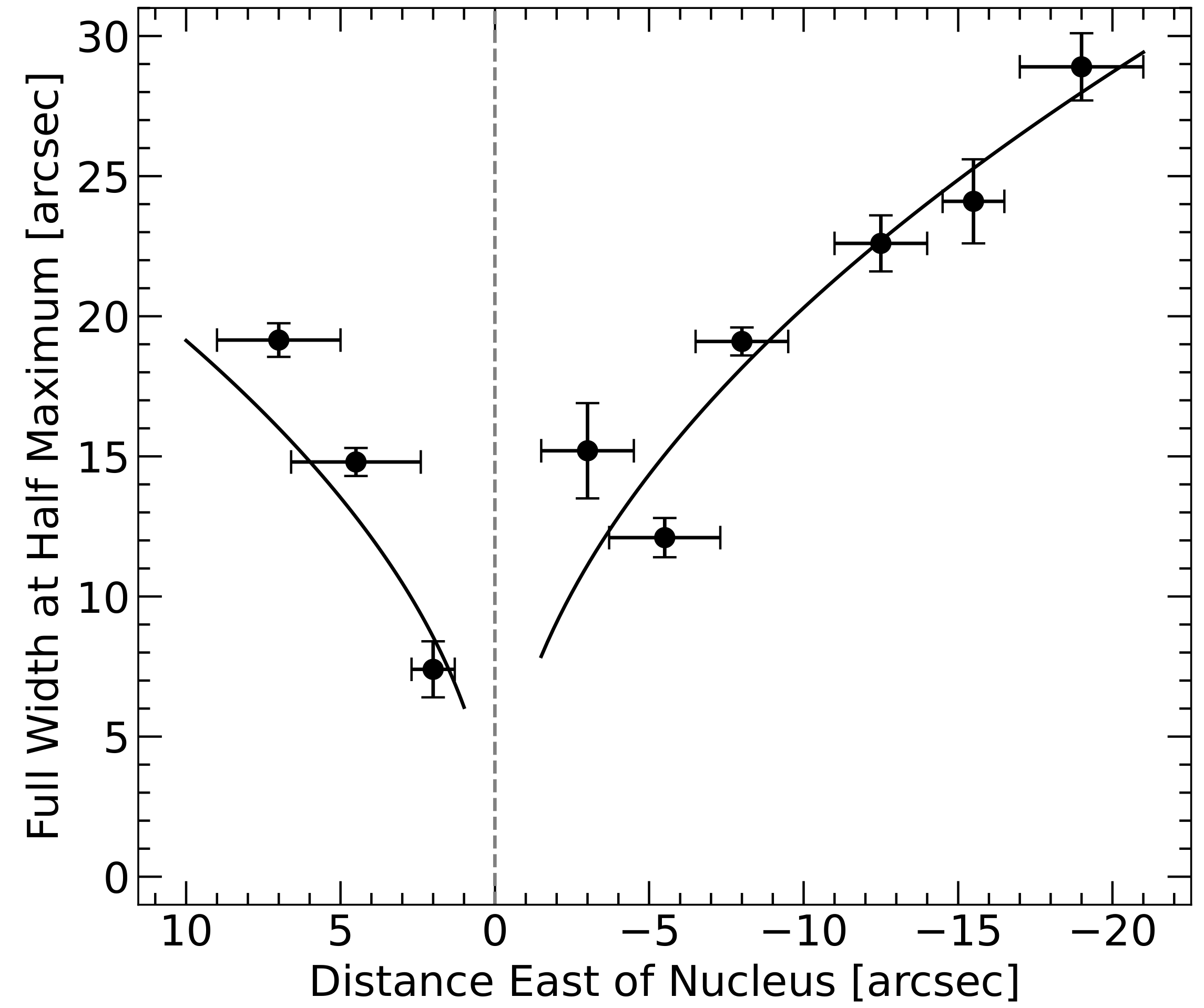}
    \caption{FWHM of the dust tail of comet E3 vs.~distance from the nucleus in the observation image taken on 10 October 2022 when Earth was located almost in the orbital plane of comet E3. The horizontal error bars show the distance range along the tail that was used to measure the FWHM, and the vertical bars indicate the range of uncertainties of the FWHM profiles. The red and blue lines denote the best-fitting relation $\theta_{\perp} \propto l^{1/2}$ between the FWHM of the tail and the distance from the nucleus.}
    \label{fwhm}
\end{figure}

The dust distribution perpendicular to the orbital plane of comet E3 can be determined from observation images taken on 06 July 2022 because at this time, the Earth was located almost in the orbital plane of comet E3 (Table \ref{geometry}). The tail width $\theta_{\perp}$ is shown along the distance from the nucleus in Figure \ref{fwhm}, where a broad tail is displayed, with $\theta_{\perp} < 28\arcsec$ up to $20\arcsec$ east of the nucleus and $\theta_{\perp} < 19\arcsec$ up to 10\arcsec west of the nucleus. The ejection speed perpendicular to the orbital plane $v_{\perp}$ was derived from the tail width $\theta_{\perp}$ according to \citet{jewitt2015episodic}
\begin{equation}
v_{\perp}=\left[\frac{\beta g_{\odot}(1)}{8 \ell R^{2}}\right]^{1 / 2} \theta_{\perp},
        \label{PerVel}
\end{equation}
%where $\beta\sim0.3$ is related to the mean radius of the dust grains $\overline{a}\sim30$ \SI{}{\um}, by $\beta = 1/a_{\SI{}{\um}}$. 
%where the value of $\beta$ is estimated as about 0.3 by $\beta = 1/a_{\SI{}{\um}}$ for the mean radius of the dust grains $\overline{a}\sim30$ \SI{}{\um}.}
where $g_{\odot}(1) = 0.006$ m s$^{-2}$ is the solar gravitational acceleration at a heliocentric distance of 1 AU, and $\ell$ is the distance from the nucleus. The measured tail width was fit to Equation (\ref{PerVel}), as shown in Figure \ref{fwhm}, and the result shows that the ejection speed is approximately $v_{\perp}\sim 89.8\beta^{1/2}$ m s$^{-1}$ on the east side of the nucleus. It is approximately $v_{\perp} \sim 83.1\beta^{1/2}$ m s$^{-1}$ on the west side of the nucleus. This $\beta^{1/2}$ dependence (i.e.,~$a^{-1/2}$ dependence) is basically consistent with that of the emission driven by gas \citep{kim2020coma}. By substituting the estimated values of the mean grain radius of $a\sim100$ \SI{}{\um} inferred from Section \ref{Dust properties} into the aforementioned fitting formula, the ejection speed perpendicular to the orbital plane
is obtained to be $v_{\perp} \sim 9$ m s$^{-1} $. The result is approximately consistent with $v_{\mathrm{ej}}\sim 14$ m s$^{-1}$ 
for the grain radius of $a\sim100$ \SI{}{\um} inferred from Section \ref{morphology}.
%The dependence of the ejection velocity on particle size, $v_{\perp} \sim \beta^{1/2}$ (i.e.~$v_{\perp}\propto a^{-1/2}$)}, 
\subsection{Sublimating area}
\label{Sublimating area}

Comet E3 is continuously active, and its activity level increases as the heliocentric distance decreases, suggesting that the sublimation of ice is likely the origin of the activity. Following \citet{hui2019c}, the mass flux of the sublimation of ice, denoted $f_s(T)$, can be calculated using the energy balance equation,
\begin{equation}
\frac{S_{\odot}}{R^{2}}(1-A)=\chi\left[\epsilon \sigma T\left(R\right)^{4}+H_{s} f_{s}(T)\right],
        \label{waterRate}
\end{equation}
where $S_{\odot}=1361$ W m$^{-2}$ is the solar constant, $\sigma = 5.67\times10^{-8}$ W m$^{-2}$ K$^{-4}$ is the standard value of the Stefan–Boltzmann constant, $A = 0.04$ is the assumed bond albedo, and $H_s (T)$ is the latent heat of volatile substances at a specific temperature, $T$. The distribution coefficients of the incident heat and the effective emissivity are assumed to be $\chi = 2$ and $\epsilon = 0.9$, respectively.
The equilibrium temperature of comet E3 varies in the range of about 162 K < $T$ < 202 K when the heliocentric distance is between 1.83 AU and 2.86 AU. Estimated from the Equation (\ref{waterRate}), the mass flux of
the H$_2$O sublimation varies approximately in the range of $1.68\times10^{-5}$ kg m$^{-2}$ s$^{-1}$ (2.86 AU) < $f_s$ < $4.47\times10^{-5}$ kg m$^{-2}$ s$^{-1}$ (1.83 AU).

The OH number production rate $Q_\mathrm{OH}$ of comet E3 is approximately $(1.16\pm0.23)\times10^{28}$ s$^{-1}$ in October 2022 \citep{jehin2022trappist}, 
from which the number production rate of water for comet E3 is estimated as $Q_\mathrm{H_2O} = (1.23\pm0.24)\times 10^{28}$ s$^{-1}$. As a result, the
mass production rate of water is derived to be $\dot{M}_{\mathrm{water}}\sim (368\pm72)$ kg s$^{-1}$ in October 2022. Assuming that the nucleus of comet E3 is a homogeneous sphere, the lower limit of the size of this nucleus that can maintain the observed sublimation activity is estimated according to \citep{jewitt2019initial}
\begin{equation}
r_{\mathrm{n}} \gtrsim \sqrt{\frac{\dot{M}_{\mathrm{water}}}{4\pi f_{\mathrm{s}}}}.
        \label{MasswaterRate}
\end{equation}
By substituting the estimated values of $\dot{M}_{\mathrm{water}}$ and $f_s$ (for H$_2$O) into Equation (\ref{MasswaterRate}), the lower limit of the size of the nucleus is determined to be $\sim(0.81\pm0.07)$ km, 
%with the corresponding sublimating area measuring approximately $5.5\times10^6$ m$^2$.
which corresponds to a sublimating area of about $8.2\times10^6$ m$^2$. In comparison, 2I/2019 Q4 (Borisov), which is classified as an interstellar comet, was activated before reaching its perihelion due to the sublimation of volatile substances, and \citep{jewitt2019initial} analyzed the data of 2I obtained with a similar observation geometry and estimated its sublimating area to be approximately $1.5\times10^6$ m$^2$.
\begin{figure}
\centering
\includegraphics[width=1\columnwidth]{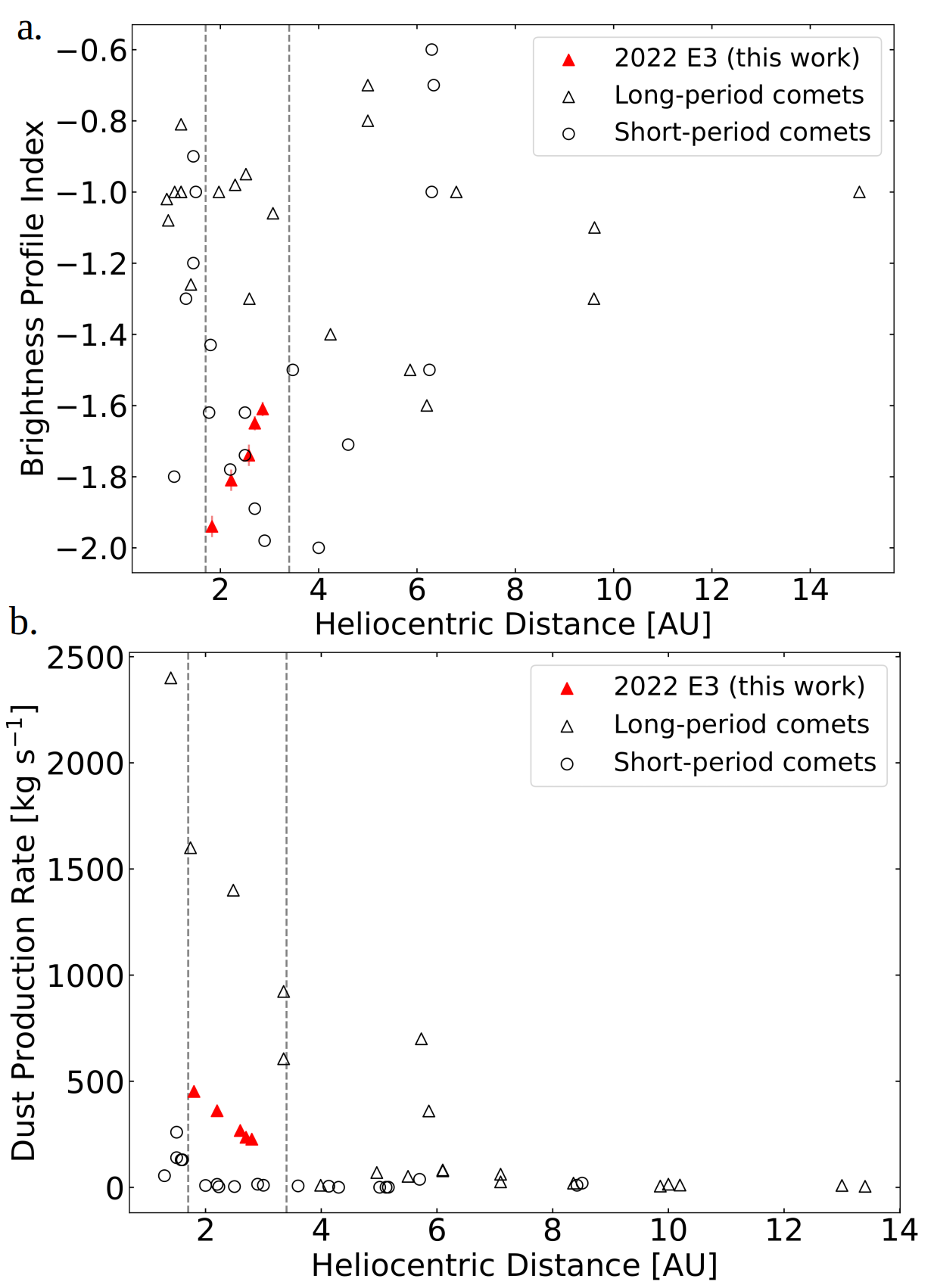}
    \caption{Comparison of the activity characteristics of comet E3 (see Section \ref{morphology} and Section \ref{Nucleus size and dust production rates}) with those of comets belonging to different types (see Appendix \ref{appdA}). a) Brightness index vs.~heliocentric distance of comet E3, long-period comets, and short-period comets. b) Dust production rates vs.~the heliocentric distance of comet E3, long-period comets, and short-period comets. The error bars of the E3 production rates are obscured by the data points. The two vertical dashed lines represent the heliocentric distances of 1.7 AU and 3.4 AU, respectively, within the range of which the observations of comet E3 that were used for analysis in this study were taken (see Table \ref{geometry}). It should be noted that the data of the listed brightness profiles and dust production rates are from two different literature collections.}
    \label{figureCompare}
\end{figure}
\subsection{Particular characteristics of E3 compared with other comets}
\label{The Particular Characteristics of E3 Compared with Other Comets}
In this section, we present a comparative analysis of the characteristics of comet E3 with those of comets belonging to different types. Figure \ref{figureCompare}a illustrates the brightness profiles of different types of comets, including 11 short-period comets, %\citep{lejoly2022radial,meech1997rotation,bolin2021initial,farnham2002mcdonald,xu2022physical,farnham2005physical,rousselot2008174p,shi2019research,shi2023monitoring},
14 long-period comets %\citep{jewitt2019distant,ivanova2021observations,ivanova2023quasi,ivanova2015ccd,betzler2023photometric,betzler2017analysis,kulyk2021optical,li2015disappearance,jewitt2019distant,opitom2015trappist,shi2019research,das2013polarimetric,lara2001activity,reyniers2009rotation,betzler2018photometric,schulz2002dust}
, and comet E3. Figure \ref{figureCompare}a clearly shows that the brightness indices of the majority of the long-period comets are higher than -1.5, whereas the brightness indices of the short-period comets are more widely distributed, and their heliocentric distances during outbursts are smaller than 7 AU. Notably, when the range of the heliocentric distance is narrowed to about [1.7, 3.4] AU (i.e.,~where the observations of comet E3 that were used for analysis in this study were taken (see Table \ref{geometry})), it becomes evident that the indices of the short-period comets are close to those of the brightness profiles of comet E3.

Figure \ref{figureCompare}b depicts the dust mass production rates for eight short-period comets%\citep{lamy2002nucleus,moreno2012comet,moreno2016dust,soja2015characteristics,clements2021dust29P,yang2019comet,ishiguro20102007,lamy1998nucleus,rousselot2008174p}
, six long-period comets%\citep{de2014herschel,korsun2008c,shi2014ccd,rousselot2014monitoring,epifani2016photometry,tricarico2014delivery,woodward2021coma}
, and comet E3, the data of which are from a different literature collection from the brightness profiles shown in Figure \ref{figureCompare}a.
When we compare the dust production rates of comets within the heliocentric distance range of about [1.7, 3.4] AU, it becomes evident that the activity level of comet E3 is significantly lower than those of other long-period comets. Furthermore, the values of the dust production rates of comet E3 at different heliocentric distances are closer to those of the short-period comets. Thus, when we correlate the activity level of comets with their heliocentric distances, as indicated in Figure \ref{figureCompare}a and \ref{figureCompare}b, and without considering other factors that might influence the activity level, such as the nucleus size, comet E3 exhibits a duality.
In other words, while the orbital characteristics of E3 classify it as a long-period comet, its activity profile is more closely aligned with those of short-period comets within the heliocentric distance range where the observation images of comet E3 that were used for the analysis in this study were taken.

\section{Summary}
\label{summary}
This paper analyzed the activity of the long-period comet 2022 E3 (ZTF) at five epochs between 13 July 2022 and 10 October 2022. The results are listed below.
 \begin{enumerate}
\item  We showed that the brightness distribution indices of the inner coma are lower than -1.5, suggesting that the particles from comet E3 are released continuously and the process of dust emission is in nonsteady state. The increase in the dust production rate with the decreasing heliocentric distance might be one reason for the nonsteady-state nature of the dust emission process.

\item The syndyne fitting results show that dust particles larger than about 10 µm contribute significantly to the observed tail. The model simulations of the 10 October 2022 image show that the radii of the majority of the large particles lingering near the nucleus range from 0.1 mm to 1 mm.

\item The radius range of the nucleus of comet E3 is estimated as between $0.81\pm0.07$ km and $2.79\pm0.01$ km, with an assumed albedo of 0.1. The dust production rates increased continuously from $241\pm3$ kg s$^{-1}$ in July 2022 to $476\pm9$ kg s$^{-1}$ in October 2022.

\item The dependence of the ejection velocity on the grain size is basically consistent with that of emission driven by gas. The ejection speed perpendicular to the orbital plane of the particle with an average radius of 100 \SI{}{\um} is estimated as $\sim9$ m s$^{-1}$.

\item The estimated water production rate is $\sim 368\pm72$ kg s$^{-1}$. The lower limit of the equilibrium sublimating area that can maintain the observed sublimation activity is approximately $8.2\times10^6$ m$^2$, which corresponds to a sphere with a radius of $0.81\pm0.07$ km.

\item Within the heliocentric distance range of about [1.7, 3.4] AU where the observation images of comet E3 that were used for the analysis in this study were taken, the following phenomena are found. The brightness indices of comet E3 are close to those of short-period comets. In addition, the values of the dust production rates of comet E3 are lower than those of long-period comets, but they agree well with those of short-period comets.
\end{enumerate}

\begin{acknowledgements}
This work was supported by the National Natural Science Foundation of China (No.~12002397, 12311530055 and 62388101), and the Shenzhen Science and Technology Program (Grant No.~ZDSYS20210623091808026). We thank Shangfei Liu for helpful discussions and suggestions. The data used in this work is generated as detailed in the text and will be shared on reasonable request to the corresponding author.
\end{acknowledgements}
%%%%%%%%%%%%%%%%%%%%%%%%%%%%%%%%%%%%%%%%%%%%%%%%%%.

\bibliography{aanda}{}
\bibliographystyle{aa}

\begin{appendix} 
\section{\mbox{Information about the comets in Figure 8}} 
% 附录的章节标题
\label{appdA}
\begin{table}[!htbp]
\captionsetup{singlelinecheck=false, justification=raggedright}
    \centering
    \renewcommand{\arraystretch}{1.3}
    \caption{\mbox{Activity characteristics of the comets}}
    \label{appen}
    \setlength{\tabcolsep}{0pt}
    \begin{tabular}{lcccc}
                \toprule
Name & Type$^{a}$ & $q^{b}$ ($R^c$) & $\dot{M}^{d}$ ($R^c$)  & References \\
\hline
17P & S & - & 2.8 (2.2), 5.4 (4.1), 0.6 (5.0), 0.8 (5.1), 0.6 (5.1),& [1]  \\
19P & S & -1.2 (1.6), -1.3 (1.4) & - & [2]  \\
22P & S  & - & 130.0 (1.6), 260.0 (1.6) & [3], [4]  \\
29P & S & - & 38.0 (5.8) & [5]  \\
45P & S & -1.8 (0.5)& - & [6] \\
46P & S & -1.1 (1.1), -1.6 (2.5), -1.4 (1.8) & 4.0 (2.5) & [6], [7], [8]  \\
60P & S & -1.6 (1.7) & - & [9]  \\
64P & S  & -0.9 (1.6) & - & [10]  \\
66P & S & - & 55.0 (1.3) & [11]  \\
67P & S &  -2.0 (2.9), -1.9 (2.7)& 15.0 (2.9), 0.5 (4.3) & [12], [13], [14]  \\
67P & S & -1.7 (2.5), -1.8 (2.2)  &  10.0 (3.0), 7 (3.6) & [12], [13], [14]  \\
81P & S  & -1.0 (1.7), -1.3 (1.7) & - & [15]  \\
174P & S  & -1.0 (6.3) & 20.0 (8.5), 10 (8.4) & [16], [17]  \\
228P & S & -1.5 (3.5) & - & [18]  \\
2019 LD$_2$ & S & -1.7 (4.6) & - & [19] \\
C/1996 Q1 & L & -1.0 (1.2) & - & [20]  \\
C/1999 S4 & L & -1.0 (0.9) & - & [21]  \\
C/2002 VQ94& L & - & 20.0 (8.4), 6.0 (9.9), 5.0 (13.4) & [22]  \\
C/2004 Q2 & L & -0.8 (1.2), -1.0 (1.2) & - & [23]  \\
C/2006 S3 & L & -1.5 (5.9) & 360.0 (5.9), 82.0 (6.1), 9.0 (13.9), 52.0 (5.6), 11.0 (10.3) & [17], [24]  \\
C/2006 S3 & L & - & 15.0 (10.0), 26.0 (7.2), 78.0 (6.1), 62.0 (7.2) &  [9], [24]  \\
C/2006 OF2 & L & -1.4 (4.2) & - & [25]  \\
C/2006 W3& L & - & 70.0 (5.0), 676.0 (3.3), 923 (3.3), 606 (3.3) & [26]  \\
C/2009 P1& L & -1.0 (2.0) & 700.0 (5.7), 1400.0 (2.5) & [27], [28]  \\
C/2010 X1& L & -1.6 (2.9) & - & [29]  \\
C/2010 S1& L & -0.7 (5.0), -1.6 (6.2) & - & [30]  \\
C/2011 KP36& L & -0.8 (5.0) & - & [31]  \\
C/2012 F6& L & -1.3 (1.9) & - & [32]  \\
C/2014 B1& L & -1.1 (9.6), -1.3 (9.6) & - & [33], [34]  \\
C/2014 S2& L & -1.0 (2.3) & - & [35]  \\
C/2013 A1& L & - & 10.0 (3.8) & [36]  \\
C/2013 US10& L & - & 1600.0 (1.7), 2400.0 (1.4) & [37]  \\
C/2017 K2& L & -1.0 (15.9), -1.0 (13.8) & - & [38]  \\
% C/2021 A1& L & - & 800.0 (2.7) & [39]  \\

\hline
\multicolumn{5}{p{19cm}}{$^a$ S: short-period comet; L: long-period comet. $^b$ Brightness index rounded to one decimal place. $^c$ Heliocentric distance rounded to one decimal place, in AU. $^d$ Dust production rate rounded to one decimal place, in kg/s. 

References: [1] \citet{ishiguro20102007}; [2] \citet{farnham2002mcdonald}; [3] \citet{lamy2002nucleus}; [4] \citet{moreno2012comet}; [5] \citet{clements2021dust29P}; [6] \citet{lejoly2022radial}; [7] \citet{meech1997rotation}; [8] \citet{lamy1998nucleus}; [9] \citet{shi2023monitoring}; [10] \citet{xu2022physical}; [11] \citet{yang2019comet}; [12] \citet{moreno2016dust}; [13] \citet{soja2015characteristics}; [14] \citet{guilbert2014pre}; [15] \citet{farnham2005physical};[16] \citet{rousselot2021new};[17] \citet{rousselot2016long}; [18] \citet{shi2014ccd}; [19] \citet{bolin2021initial}; [20] \citet{guilbert2014pre}; [21] \cite{schulz2002dust}; [22] \citet{korsun2014distant}; [23] \citet{reyniers2009rotation}; [24] \citet{rousselot2014monitoring}; [25] \citet{kulyk2021optical}; [26] \citet{de2014herschel}; [27] \citet{epifani2016photometry}; [28] \citet{bodewits2014evolving}; [29] \citet{li2015disappearance}; [30] \citet{ivanova2015ccd}; [31] \citet{ivanova2021observations}; [32] \citet{opitom2015trappist}; [33] \citet{jewitt2019discus}; [34] \citet{ivanova2023quasi}; [35] \citet{betzler2023photometric}; [36] \citet{tricarico2014delivery}; [37] \citet{woodward2021coma}; [38] \citet{jewitt2019distant}.}
         \end{tabular}
    \end{table}
\end{appendix}

\end{document}